\documentclass[12pt]{article}

\begin{document}

\title{\bf Two-Loop Superstrings in Hyperelliptic Language II:
 the Cosmological Constant and the
Non-Renormalization Theorem}

\author{Zhu-Jun Zheng\thanks{Supported in part by Math. Tianyuan Fund
with grant Number 10226002 and the Natural Science Foundation of
Educational Committee of Henan Province with grant Number
2000110010.  }\\ \\
Institute of Mathematics, Henan University \\
Kaifeng 475001, P. R. China\\ and\\
Institute of Theoretical Physics,
Chinese Academy of Sciences\\
P. O. Box 2735,  Beijing 100080, P. R. China \\  \\
Jun-Bao Wu \\School of Physics, Peking University \\
Beijing 100871, P. R. China\\ \\
Chuan-Jie Zhu\thanks{Supported in
part by fund from the National Natural Science Foundation of China
with grant Number
90103004.} \\
Institute of Theoretical Physics,
Chinese Academy of Sciences\\
P. O. Box 2735,  Beijing 100080, P. R. China}

\maketitle
\newpage
\begin{abstract}
The vanishing of the cosmological constant and the
non-renormali-zation theorem are verified at two loops by explicit
computation using the hyperelliptic language and the newly
obtained chiral measure of D'Hoker and Phong. A set of identities
is found which is  used in the verification of the
non-renormalization theorem and leads to a great simplification of
the calculation of the four-particle amplitude at two loops.
\end{abstract}


\section{Introduction}

Although we believe that superstring theory is finite in
perturbation at any order \cite{GreenSchwarz1, GreenSchwarz2,
GreenSchwarz3, Martinec}, a rigorous proof is still lacking
despite great advances in the covariant formulation of superstring
perturbation theory \'a la Polyakov. In particular, there is a
non-renormalization theorem \cite{Martinec}. In spite of the
efforts of many authors, it is very difficult to verify this
theorem explicitly. Even in the case of the cosmological constant,
i.e. the vacuum amplitude, this problem has not been completely
solved. At two loops these problems were solved explicitly by
using the hyperelliptic formalism in a series of papers
\cite{GavaIengoSotkov, IengoZhu1, Zhu, IengoZhu2, IengoZhu3}. The
explicit result was also used by Iengo \cite{Iengo} to prove the
vanishing of perturbative correction to the $R^4$ term
\cite{GrossWitten} at two loops, in agreement with the indirect
argument of Green and Gutperle \cite{GreenGutperle}, Green,
Gutperle and Vanhove \cite{Green2}, and Green and Sethi
\cite{GreenSethi} that the $R^4$ term does not receive
perturbative contributions beyond one loop. Recently, Stieberger
and Taylor \cite{Stieberger}  also used the result of
\cite{IengoZhu2} to prove the vanishing of the heterotic two-loop
$F^4$ term. For some closely related works we refer the reader to
the reviews \cite{Green3, Kiritsis}. In the general case, there is
no satisfactory solution. For a review of these problem we refer
the reader to \cite{DHokerPhong1, DHokerPhong6}.

Recently two-loop superstring was studied by D'Hoker and Phong. In
a series of papers \cite{DHokerPhong2, DHokerPhong3, DHokerPhong4,
DHokerPhong5} (for a recent review see \cite{DHokerPhong6}),
D'Hoker and Phong found an unambiguous and slice-independent
two-loop superstring measure on moduli space for even spin
structure from first principles.

Although their result is quite explicit, it is still a difficult
problem to use it in actual computation. In \cite{DHokerPhong5},
D'Hoker and Phong used their result to compute explicitly the
chiral measure by choosing the split gauge and proved the
vanishing of the cosmological constant and the non-renormalization
theorem \cite{DHokerPhong7, Martinec}. They also computed the
four-particle amplitude in another forthcoming paper
\cite{DHokerPhong8}. Although the final results are exactly the
expected, their computation is quite difficult to follow because
of the use of theta functions.\footnote{In \cite{Lechtenfeld5},
the two-loop 4-particle amplitude was also computed by using theta
functions. Its relation with the previous explicit result
\cite{IengoZhu2} has not been clarified.}  Also modular invariance
is absurd in their computations because of the complicated
dependence between the 2 insertion points (the insertion points
are also spin structure dependent).

Although the vanishing of the cosmological constant and the
non-renorma-lization theorem was proved explicitly in previous
works \cite{GavaIengoSotkov, IengoZhu1, Zhu}, it would be
interesting to study this problem again by using the newly
obtained result of D'Hoker and Phong. The main purpose of this
study is as a warm up exercise for the computation of the possibly
non-vanishing four-particle amplitude. As we will see in this
paper, some expressions are non-vanishing after summation over
spin structures. Nevertheless the combination of the symmetry of
the computed expression and the relevant kinematic factor gives a
vanishing result. In a previous paper \cite{AllZhu1}, we report
the main results of our computation of two loop superstring theory
by using hyperelliptic language. In this paper we will present the
details for the proof of the vanishing of the cosmological
constant and the non-renormalization theorem. The computation of
the non-vanishing four-particle amplitude is given in another
publication \cite{AllZhu3}.

The organization of this paper is as follows. In the next section
we will recall the relevant results of hyperelliptic
representation of the genus 2 Riemann surface and set our
notations for all the correlators. In section 3 we recall the
results of D'Hoker and Phong for the chiral measure. In section 4
we computed explicitly all the relevant quantities in the chiral
measure. Here we mainly concentrated on the spin structure
dependent parts.  In section 5 we established a set of identities
and proved the vanishing of the cosmological constant. The
identities will also be used in the next section in the
verification of the non-renormalization theorem. Here modular
invariance is maintained explicitly. In this section we also
discuss the importance of taking the limit $\tilde p_1 \to
q_{1,2}$ and mention the (six) Riemann identities which are not
fully modular invariant. In section 6 we proved the
non-renormalization theorem. In particular we study carefully the
most difficult part of the three-particle amplitude. Here the
symmetry of the relevant kinematic factor is very important in the
proof of the non-renormalization theorem. The (point-wise)
vanishing of all the 1-, 2- and 3-particle amplitude leads a great
simplification of the computation of the 4-particle amplitude
\cite{AllZhu3}.

Here we   note again that D'Hoker and Phong have  proved that the
cosmological constant and the 1-, 2- and 3-point functions are
zero point-wise in moduli space \cite{DHokerPhong7}. They have
also computed the 4-particle amplitude \cite{DHokerPhong8}. The
agreement of the results from these two different gauge choices
and two different methods of computations is another proof of the
validity of the new supersymmetric gauge fixing method at two
loops.

\section{Genus 2 hyperelliptic Riemann surface}

First we remind that a genus-g Riemann surface, which is the
appropriate world sheet for one and two loops, can be described in
full generality by means of the hyperelliptic
formalism.\footnote{Some early works on two loops computation by
using hyperelliptic representation are \cite{Knizhnik, Morozov,
Morozov1, Morozov2, Lechtenfeld, Bershadsky, Moore} which is by no
means the complete list.}  This is based on a representation of
the surface as two sheet covering of the complex plane described
by the equation:
\begin{equation}
y^2(z) = \prod_{i=1}^{2g+2} ( z- a_i), \label{covering}
\end{equation}
The complex numbers $a_{i}$, $(i=1,\cdots,2g+2)$ are the $2g+2$
branch points, by going around them one passes from one sheet to
the other. For two-loop ($g=2$) three of them represent the moduli
of the genus 2 Riemann surface over which the integration is
performed, while the other three can be arbitrarily fixed. Another
parametrization of the moduli space is given by the period matrix.

At genus 2, by choosing a canonical homology basis of cycles we
have the following list of 10 even spin structures:
\begin{eqnarray}
\delta_1 \sim \left[ \begin{array}{cc} 1 & 1\\ 1 & 1 \end{array}
\right]  \sim (a_1 a_2  a_3|a_4 a_5 a_6),  & & \delta_2 \sim
\left[
\begin{array}{cc} 1 & 1\\ 0 & 0 \end{array} \right] \sim
(a_1a_2a_4|a_3a_5a_6), \nonumber
\\
\delta_3 \sim \left[ \begin{array}{cc} 1 & 0\\ 0 & 0 \end{array}
\right]  \sim (a_1a_2a_5|a_3a_4a_6),  & & \delta_4 \sim \left[
\begin{array}{cc} 1 & 0\\ 0 & 1 \end{array} \right] \sim
(a_1 a_2 a_6|a_3 a_4 a_5), \nonumber
\\
\delta_5 \sim \left[ \begin{array}{cc} 0 & 1\\ 0 & 0 \end{array}
\right]  \sim (a_1 a_3 a_4|a_2 a_5 a_6), & & \delta_6 \sim \left[
\begin{array}{cc} 0 & 0\\ 0 & 0 \end{array} \right] \sim
(a_1 a_3 a_5|a_2 a_4 a_6), \nonumber
\\
\delta_7 \sim \left[ \begin{array}{cc} 0 & 0\\ 0 & 1 \end{array}
\right]  \sim (a_1 a_3 a_6|a_2 a_4 a_5), & & \delta_8 \sim \left[
\begin{array}{cc} 0 & 0\\ 1 & 1 \end{array} \right] \sim
(a_1 a_4 a_5|a_2 a_3 a_6), \nonumber
\\
\delta_9 \sim \left[ \begin{array}{cc} 0 & 0\\ 1 & 0 \end{array}
\right]  \sim (a_1 a_4 a_6|a_2 a_3 a_5), & & \delta_{10} \sim
\left[
\begin{array}{cc} 0 & 1\\ 1 & 0 \end{array} \right] \sim
(a_1 a_5 a_6|a_2 a_3 a_4). \nonumber
\end{eqnarray}
We will denote an even spin structure as $(A_1 A_2 A_3|B_1 B_2
B_3)$. By convention $A_1= a_1$. For each even spin structure we
have a spin structure dependent factor from determinants which is
given as follows \cite{GavaIengoSotkov}:
\begin{equation}
Q_\delta = \prod_{i <j} (A_i-A_j)(B_i-B_j).
\end{equation}
This is a degree 6 homogeneous polynomials in $a_i$.

At two loops there are two odd supermoduli and this gives two
insertions of supercurrent  at two different points $x_1$ and
$x_2$.  Previously the chiral measure was derived in
\cite{Verlinde, DHokerPhong1} by a simple projection from the
supermoduli space to the even moduli space. This projection does't
preserve supersymmetry and there is a residual dependence on the
two insertion points. This formalism was used in
\cite{GavaIengoSotkov, IengoZhu1, Zhu, IengoZhu2}. In these papers
we found that it is quite convenient to choose these two insertion
points as the two zeros of a holomorphic abelian differential
which are moduli independent points on the Riemann surface. In
hyperelliptic language these two points are the same points on the
upper and lower sheet of the surface. We denote these two points
as $x_1=x+$ (on the upper sheet) and $x_2=x-$ (on the lower
sheet). We made these convenient choices again in \cite{AllZhu1}
and will make the same choices in this paper and \cite{AllZhu3}.

In the following we will give some formulas in hyperelliptic
representation which will be used later. First all the relevant
correlators are given by\footnote{We follow closely the notation
of \cite{DHokerPhong3}. }
\begin{eqnarray}
\langle \psi^\mu   (z) \psi^\nu   (w) \rangle & = &
-\delta^{\mu\nu} G_{1/2}[\delta] (z,w) = - \delta^{\mu\nu}S_\delta
(z,w),
  \\
\langle \partial_z X^\mu  (z) \partial_w X^\nu  (w) \rangle & = &
-\delta^{\mu\nu}\partial_z
\partial_w \ln E(z,w),
 \\
\langle b(z) c(w) \rangle &=& +G_2 (z,w), \\
\langle \beta (z) \gamma (w) \rangle &=& -G_{3/2}[\delta] (z,w),
\end{eqnarray}
where
\begin{eqnarray}
& & S_{\delta}(z,w) = { 1\over z-w} \, { u(z) + u(w) \over 2
\sqrt{u(z) u(w) } } , \\
& & u(z) = \prod_{i=1}^3 \left( z-A_i \over z-B_i\right)^{1/2}, \\
& & G_2(z,w) = -H(w,z) + \sum_{a=1}^3  H(w,p_1) \, \varpi_a(z,z),
\\
& & H(w,z) = { 1\over 2(w- z)} \,\left( 1 + { y(w) \over
y(z) }\right) \, { y(w) \over y(z) }, \\
& & G_{3/2}[\delta](z,w) = - P(w,z) + P(w,q_1) \psi_1^*(z) +
P(w,q_2)\psi_2^*(z), \label{eq51} \\
& & P(w,z) = {1\over \Omega(w)}\, S_{\delta}(w,z)\Omega(z),
\end{eqnarray}
where $\Omega(z)$ is an abelian differential satisfying
$\Omega(q_{1,2}) \neq 0$ and otherwise arbitrary. These
correlators were adapted from \cite{Iengo2}. $\varpi_a(z,w)$ are
defined in \cite{DHokerPhong2} and $\psi^*_{1,2}(z)$ are the two
holomorphic $3\over 2$-differentials. When no confusion is
possible, the dependence on the spin structure $[\delta]$ will not
be exhibited.

In order take the limit of $x_{1,2}\to q_{1,2}$ we need the following
expansions:
\begin{eqnarray}
G_{3/2} (x_2, x_1) &=& {1 \over x_1 - q_1} \psi ^* _1 (x_2)
      - \psi ^* _1 (x_2) f_{3/2} ^{(1)} (x_2) +O(x_1 - q_1),
\\
G_{3/2} (x_1, x_2) &=& {1 \over x_2 - q_2} \psi ^* _2 (x_1)
      - \psi ^* _2 (x_1) f_{3/2} ^{(2)} (x_1)  +O(x_2 - q_2),
\end{eqnarray}
for $x_{1,2}  \to q_{1,2}$. By using the explicit expression of
$G_{3/2}$ in (\ref{eq51}) we have
\begin{eqnarray}
f_{3/2} ^{(1)} (q_2) & = & - {\partial_{q_2} S(q_1,q_2) \over
S(q_1,q_2)
} + \partial\psi^*_2(q_2), \label{eq54}\\
f_{3/2} ^{(2)} (q_1) & = &   {\partial_{q_1} S(q_2,q_1) \over
S(q_1,q_2) } + \partial\psi^*_1(q_1) = f_{3/2} ^{(1)}(q_2)|_{ q_1
\leftrightarrow q_2 } . \label{eq55}
\end{eqnarray}

The quantity $\psi^*_\alpha (z)$'s are holomorphic $3\over
2$-differentials and are constructed as follows:
\begin{equation}
\psi^*_\alpha (z) = (z-q_\alpha)S(z,q_\alpha)  \,
{y(q_\alpha)\over y(z)} \, , \qquad \alpha = 1, 2.
\end{equation}
For $z=q_{1,2}$ we have
\begin{eqnarray} &  & \psi^*_\alpha (q_\beta ) =
\delta_{\alpha,\beta}, \\
& & \partial \psi^*_1 (q_2) = -\partial \psi^*_2 (q_1) =
S(q_1,q_2) = {i\over 4}S_1(q), \\
& & \partial \psi^*_1 (q_1) =  \partial \psi^*_2 (q_2) =
- {1\over 2} \Delta_1(q),  \\
& & \partial^2  \psi^*_1 (q_1) =  \partial^2 \psi^*_2 (q_2) =
{1\over 16}S_1^2(q)  + {1\over 4}\Delta_1^2(q) + {1\over
2}\Delta_2(q),
\end{eqnarray}
where
\begin{eqnarray}
\Delta_n(x) & \equiv & \sum_{i=1}^6 {
1\over (x - a_i)^n }, \\
S_n(x) & \equiv &   \sum_{i=1}^3 \left[ { 1\over (x - A_i)^n } - {
1\over (x - B_i)^n }\right],
\end{eqnarray}
for $  n = 1, 2$. This shows that $\partial\psi^*_\alpha(q_{\alpha
+1})$ and $\partial^2\psi^*_\alpha(q_{\alpha})$ are spin structure
dependent.

\section{The chiral measure: the result of D'Hoker and Phong}

The chiral measure obtained in \cite{DHokerPhong2, DHokerPhong3,
DHokerPhong4, DHokerPhong5} after making the choice $x_\alpha =
q_\alpha$ ($\alpha= 1, 2$) is
\begin{eqnarray}
{\cal A} [\delta] & = & i {\cal Z} \biggl \{ 1  + {\cal X}_1 + {\cal
X}_2 + {\cal X}_3 + {\cal X}_4 +  {\cal X}_5 + {\cal X}_6 \biggr
\},
\nonumber \\
{\cal Z} & = & {\langle  \prod _a b(p_a) \prod _\alpha \delta (\beta
(q_\alpha)) \rangle \over \det \omega _I \omega _J (p_a) } ,
\end{eqnarray}
and the ${\cal X}_i$ are given by:
\begin{eqnarray}
{\cal X}_1 + {\cal X}_6 &=& {\zeta ^1 \zeta ^2 \over 16 \pi ^2}
\biggl [ -\langle \psi(q_1)\cdot \partial X(q_1) \, \psi(q_2)\cdot
\partial X(q_2) \rangle  \nonumber  \\
&& \hskip -1cm
 - \partial_{q_1} G_2 (q_1,q_2) \partial\psi^*_1 (q_2)
 + \partial_{q_2} G_2 (q_2,q_1) \partial\psi^*_2 (q_1)
\nonumber \\
&& \hskip -1cm + 2   G_2 (q_1,q_2) \partial\psi^*_1 (q_2)  f_{3/2}
^{(1)} (q_2) - 2   G_2 (q_2,q_1) \partial\psi^*_2 (q_1)  f_{3/2}
^{(2)} (q_1) \biggr ] \, ,
 \\
{\cal X}_2 + {\cal X}_3 &=&  {\zeta ^1 \zeta ^2 \over 8 \pi ^2}
S_\delta (q_1,q_2) \nonumber \\
&& \hskip  1cm  \times \sum_{a=1}^3 \tilde\varpi_a  (q_1, q_2)
\biggl [ \langle T(\tilde p_a)\rangle + \tilde B_2(\tilde p_a) +
\tilde B_{3/2}(\tilde p_a) \biggr ]\, , \label{eq65}  \\
{\cal X}_4 &=& {\zeta ^1 \zeta ^2 \over 8 \pi ^2} S_\delta
(q_1,q_2) \sum _{a=1}^3 \biggl [ \partial_{p_a} \partial_{q_1} \ln
E(p_a,q_1) \varpi^*_a(q_2) \nonumber \\
& & \hskip  1cm+ \partial_{p_a}
\partial_{q_2} \ln E(p_a,q_2) \varpi ^*_a(q_1) \biggr ]\, ,
 \\
{\cal X}_5 &=& {\zeta ^1 \zeta ^2 \over 16 \pi ^2} \sum_{a=1}^3
\biggl
[ S_\delta (p_a, q_1) \partial_{p_a} S_\delta (p_a,q_2) \nonumber \\
& & \hskip  1cm - S_\delta (p_a, q_2) \partial_{p_a} S_\delta
(p_a,q_1) \biggr ] \varpi_a (q_1,q_2) \, .
\end{eqnarray}
Furthermore, $\tilde B_2$ and $\tilde B_{3/2}$ are given by
\begin{eqnarray}
\tilde B_2(w) & = & -2 \sum _{a=1}^3 \partial_{p_a} \partial_w \ln
E(p_a,w) \varpi^*_a (w) \, , \\
\tilde B_{3/2}(w) &=& \sum_{\alpha=1}^2  \biggr(G_2 (w,q_\alpha)
\partial_{q_\alpha} \psi^*_\alpha (q_\alpha) + {3 \over 2}
\partial_{q_\alpha}
G_2 (w,q_\alpha) \psi^*_\alpha (q_\alpha) \biggr)  \,  .
\end{eqnarray}
In comparing with \cite{DHokerPhong4} we have written ${\cal
X}_2$, ${\cal X}_3$ together and we didn't split $T(w)$ into
different contributions. We also note that in eq.~(\ref{eq65}) the
three arbitrary points $\tilde p_a$ ($a=1,2,3$) can be different
from the three insertion points $p_a$'s of the $b$ ghost field.
The symbol $\tilde\varpi_a$ is obtained from $\varpi_a$ by
changing $p_a$'s to $\tilde p_a$'s. In the following computation
we will take the limit of $\tilde p_1 \to q_1$. In this limit we
have $\tilde\varpi_{2,3}(q_1,q_2) = 0$ and
$\tilde\varpi_1(q_1,q_2) = -1$. This choice greatly simplifies the
formulas and also make the summation over spin structure doable
(see below and \cite{AllZhu1,AllZhu3}).

\section{The chiral measure in hyperelliptic language}

The strategy we will follow is  to isolate all the spin structure
dependent parts first. As we will show in the following the spin
structure dependent factors are just $S(q_1,q_2)$,
$\partial_{q_2}S(q_1,q_2)$ and the Szeg\"o kernel if we also
include the vertex operators. Before we do this we will first
write the chiral measure in hyperelliptic language and take the
limit of $\tilde p_1 \to q_1$.

Let's start with ${\cal X}_5$. We have
\begin{equation}
S(z,q_1)\partial_zS(z,q_2) - S(z,q_2)\partial_zS(z,q_1)  = {i
\over 4 (z-q)^2} S_1(z) .
\end{equation}
So the spin structure dependent factor from ${\cal X}_5$ is
effectively $S(z+,z-)$ as shown by the following formulas:
\begin{eqnarray}
S(q_1,q_2) & = & - S(q_2,q_1) = {i \over 4 } \, S_1(q) \, , \\
\partial_{q_2}S(q_1,q_2) & =  &  - \partial_{q_1}S(q_2,q_1)
= - {i\over 8  } S_2(q)  \, .
\end{eqnarray}

For ${\cal X}_4$, the spin structure dependent factor is simply
$S_1(q) \propto S(q_1,q_2)$ as $\ln E(p_a,q_b)$ and
$\varpi^*_a(q_b)$ are spin structure independent (their explicit
expressions are not needed in this paper and will be given in
\cite{AllZhu3}).

For ${\cal X}_2 + {\cal X}_3$, we first compute the various
contributions from the different fields. The total stress energy
tensor is:
\begin{eqnarray}
T(z) & = & - {1\over 2}: \partial_zX(z)\cdot \partial_zX(z): +
{1\over 2} :\psi(z)\cdot\partial_z\psi(z): \nonumber \\
& &  - :( \partial b c + 2 b\partial c + {1\over
2}\partial\beta\gamma + {3\over 2}\beta\partial\gamma)(z):
\nonumber \\
& \equiv & T_X(z) + T_{\psi}(z) + T_{bc}(z) + T_{\beta\gamma}(z)
\, ,
\end{eqnarray}
in an obvious notations. The various contributions are
\begin{eqnarray}
T_X(w) & = & -10 T_1(w), \\
T_{\psi}(w) & = & 5 \tilde g_{1/2}(w) =  {5 \over 32}\, (S_1(w))^2, \\
T_{bc}(w) & = & \tilde g_2(w) - 2 \partial_wf_2(w), \\
T_{\beta\gamma}(w) & = & -\tilde g_{3/2}(w) + {3\over 2}\partial_w
f_{3/2}(w),
\end{eqnarray}
where
\begin{eqnarray}
f_2(w) & = & -{3\over 4} \, \Delta_1(w) +
\sum_{a=1}^3 H(w,p_a) \varpi_a(w,w),  \\
\tilde g_2(w) & = & {5 \over 16} \Delta^2_1(w)  + {3\over 8}\,
\Delta_2(w) \nonumber \\
&  & \hskip -1.5cm + \sum_{a=1}^3 H(w,p_a) \varpi_a(w,w) \left(
{ 1\over w-p_{a+1}}+{1\over w-p_{a+2}} - \Delta_1(w) \right) ,\\
f_{3/2}(w) & = & {\Omega'(w)\over \Omega(w)} + {\Omega(q_1)\over
\Omega(w)} \, S(w,q_1)\psi^*_1(w) + {\Omega(q_2)\over \Omega(w)}
\, S(w,q_2)\psi^*_2(w), \label{eq38} \\
\tilde g_{3/2}(w) & = & {1\over 2}\,{\Omega''(w)\over \Omega(w)}
+{1\over 32}\, (S_1(w))^2 \nonumber \\
& & +{\Omega(q_1)\over \Omega(w)} \, S(w,q_1)\partial\psi^*_1(w) +
{\Omega(q_2)\over \Omega(w)} \, S(w,q_2)\partial\psi^*_2(w).
\label{eq39}
\end{eqnarray}

As we said in the last section we will take the limit of $w\to
q_1$. In this limit $T_{\beta\gamma}(w)$ is singular and we have
the following expansion:
\begin{equation}
T_{\beta\gamma}(w)    =   - { 3/2\over (w-q_1)^2} -
{\partial\psi^*_1(q_1)\over w-q_1} -{1\over8}\Delta_1^2(q)  - {
1\over 32}S_1^2(q) + O(w-q_1).
\end{equation}
The dependence on the abelian differential $\Omega(z)$ drops out.
These singular terms are cancelled by similar singular terms in
$\tilde B_{3/2}(w)$. By explicit computation we have:
\begin{eqnarray}
& & \tilde B_{3/2}(w)   =     { 3/2\over (w-q_1)^2} +
{\partial\psi^*_1(q_1)\over w-q_1}   - {1\over 4}\Delta_1^2(q) + {
3\over 4}\Delta_2(q) \nonumber \\
& & \qquad  - \left( {1\over p_1-q} \, { (q-p_2)(q-p_3) \over
(p_1-p_2)(p_1-p_3)} \, \Delta_1(q) + ... \right)
\nonumber \\
& & \qquad - { 3\over 2} \left( {1\over (p_1-q)^2} \,
{(q-p_2)(q-p_3) \over (p_1-p_2)(p_1-p_3)} + ... \right) +
O(w-q_1).
\end{eqnarray}
where $...$ indicates two other terms obtained by cyclic
permutating $(p_1,p_2,p_3)$. By using the above explicit result we
see that the combined contributions of $T_{\beta\gamma}(w)$ and
$\tilde B_{3/2}(w)$ are non-singular in the limit of $w\to q_1$.
We can then take $\tilde p_1\to q_1$ in ${\cal X}_2 + {\cal X}_3$.
In this limit only $a=1$ contributes to ${\cal X}_2+{\cal X}_3$.
This is because $\tilde\varpi_{2,3}(q_1,q_2) = 0$ and
$\tilde\varpi_1(q_1,q_2) = -1$. $T_1(w)$ and $T_{bc}(w)$ are
regular in this limit and spin structure independent. In summary,
the spin structure dependent factors from ${\cal X}_2 + {\cal
X}_3$ are the following two kinds (not including the vertex
operators which will be consider later in section 6):
\begin{equation}
S_1(q)  \propto  S(q_1,q_2), \quad \hbox{and}\quad (S_1(q))^3 .
\end{equation}

Here we note that if we don't take the limit of $w \to q_1$ (or $w
\to q_2$ which has the same effect), the spin  structure dependent
factors from ${\cal X}_2 + {\cal X}_3$ would be much more
complicated. For example we will have a factor of the following
kind:
\begin{equation}
S_1(q) (S_1(w))^2 .
\end{equation}
The summation over spin structure with this factor will give a
non-vanishing contribution as we will see later in eq.
(\ref{eq67}). We will discuss this point later in section 7.

Finally we come to ${\cal X}_1 + {\cal X}_6$. By using the
explicit results given in eqs.~(\ref{eq54})--(\ref{eq55}), we have
\begin{eqnarray} {\cal X}_1 +
{\cal X}_6 & = &    \langle  \partial X(q_1)
 \cdot\partial X(q_2)   \rangle \, S(q_1,q_2)
\nonumber \\
& & - (\partial_{q_1}G_2(q_1,q_2) +
\partial_{q_2}G_2(q_2,q_1) ) S(q_1,q_2)
\nonumber \\
& &   + 2 ( G_2(q_1,q_2) + G_2(q_2,q_1) ) \nonumber \\
& & \times  (\partial\psi_1^*(q_1) S(q_1,q_2) -
\partial_{q_2}S(q_1,q_2) )  .
\end{eqnarray}
As $G_2(q_1,q_2)$ is spin structure independent, we see that all
the spin structure dependent factors are the following two kinds:
\begin{equation}
 S(q_1,q_2) = {i\over 4} S_1(q),
\end{equation}
and
\begin{equation}
\partial_{q_2} S(q_1,q_2) = {i \over 8} \, S_2(q) .
\end{equation}
Here it is important that the factor $\partial\psi^*_1(q_2)$
cancels the factor $S(q_1,q_2)$ appearing in the denominator of
$f^{(1)}_{3/2}(q_2)$.

From all the above results we see that all the spin structure
dependent parts (for the cosmological constant) are as follows:
\begin{equation}
c_1 S_1(q) + c_2 S_2(q) + c_3 S_1^3(q)+ \sum_{a=1}^3 d_a S_1(p_a),
\label{eqform}
\end{equation}
where $c_{1,2,3}$ and $d_a$'s are independent of spin structure.
In computing the $n$-particle amplitude there are more spin
structure factors coming from the correlators of $\psi$. We will
include these terms when we discuss the non-renormalization
theorem.

\section{The vanishing of the cosmological constant}

The vanishing of the cosmological constant is proved by using the
following identities:
\begin{eqnarray}
& & \sum_\delta \eta_\delta Q_\delta S_n(x) = 0, \label{eq001} \\
& & \sum_\delta \eta_\delta Q_\delta S_1^3(x) = 0, \label{eq002}
\end{eqnarray}
for $n=1,2$ and arbitrary $x$. Let us explain these identities in
detail.

First we write down explicitly the simplest example:
\begin{equation}
M(x, a ) = \sum_\delta \eta_\delta \, \prod_{i<j}^3
(A_i-A_j)(B_i-B_j) \sum_{k=1}^3\left[ {1\over x - A_k} - {1\over x
- B_k}\right].
\end{equation}
By a Mobius transformation we have:
\begin{eqnarray}
M(x,a ) & = & { y^4(x)} \, \sum_\delta \eta_\delta \,
\prod_{i<j}^3 (\tilde A_i-\tilde A_j)(\tilde B_i-\tilde B_j)
\sum_{k=1}^3[   \tilde A_k  -   \tilde B_k ] \nonumber \\
& \equiv & {  y^4(x)} \, M(\tilde a ),
\end{eqnarray}
where $\tilde a_i = {1\over x- a_i}$.

As it was shown in \cite{GavaIengoSotkov}, there is a unique set
of phases $\eta_\delta$ for which $M(a )$ (and $M(x,a)$) is
modular invariant in the following sense: for every interchanging
$a_i \leftrightarrow a_j$ ($i\neq j$), $M(a )$ got an overall
``$-$" sign, i.e. $M(a )$ is antisymmetric for every interchange
of the branch point $a_i$'s. The phases are:
\begin{equation}
\eta_1 = -\eta_2 = \eta_3 = -\eta_4 = \eta_5 = -\eta_6 = \eta_7 =
\eta_8 = -\eta_9 = \eta_{10} = 1.
\end{equation}
It is tedious to check explicitly that $M(a)$ is indeed
antisymmetric for every interchange of the branch points by using
the above set of phases. In doing so we see quite clearly that the
factor $\sum_{k=1}^3[    A_k  -     B_k ]$ is also important
because sometimes it also gives a ``$-$" sign when we interchange
$a_1$ with other branch points.

Here we remark that eq.~(\ref{eq001}) is still true if we neglect
the factor $S_1(x)$. In fact these are exactly the Riemann
identities for the $\theta$-constants by using the Thomae formula
\cite{Fay}:
\begin{equation}
\Theta^4_\delta(0) = \pm {\rm det}^2
K \prod_{i<j}^3 A_{ij}B_{ij} .
\end{equation}
 There are 6 set of
phases which satisfies eq.~(\ref{eq001}). These correspond to the
convention of setting $A_1$ to be any of the one fixed branch
points, i.e. a choice of odd spin structures. As we can see from
the above, a Riemann identity expression is not fully modular
invariant and it is only invariant under the subgroup of modular
transformations which leaves the fixed branch point invariant,
i.e.  any interchange of $a_i \leftrightarrow a_j$ but not with
$A_1$. Even if the Riemann identities guarantees the vanishing of
the cosmological constant if we blindly neglect the extra factors
$S_1(x)$ and $S_1^3(x)$, they are not powerful enough to prove the
non-renormalization theorem, not mentioning the explicit
computation of the possibly non-vanishing 4-particle amplitude.
(See more about this point at the end of this section.)

Now we proved that $M(a)$ is indeed modular invariant, it is
trivial to prove that it is 0. The trick is as follows (which is
quite useful in what follows in the proof of non-renormalization
theorem and the calculation of the four-particle amplitude).
Because $M(a)$ is a homogeneous polynomial (of degree of 7) in
$a_i$ and it is vanishing whenever $a_i=a_j$, it should be
proportional to $P(a) \equiv \prod_{i<j}(a_i-a_j)$ which is a
homogeneous polynomial of degree 15 in $a_i$. One see immediately
that the power of $a_i$ can't be matched. So $M(a)$ must vanish.
An explicit computation by computer also verifies this
result.\footnote{Expanding $Q_\delta$ gives 36 different terms and
multiplying with $(A_1+A_2+A_3- B_1-B_2-B_3)$ gives 72 different
terms. So we have 720 terms in the sum which must cancel each
other.}

The other identities in eq.~(\ref{eq001}) and eq.~(\ref{eq002})
can be proved similarly. We note that the power 3 in eq.
(\ref{eq002}) is important to make the expression modular
invariant. In fact for all odd powers $n$, the following
expression is modular invariant:
\begin{equation}
M_{1,n}(x, a) = \sum_\delta \eta_\delta Q_\delta S_1^n(x) .
\end{equation}
By power counting we have
\begin{equation}
M_{1,n}(x,a) = 0, \qquad \hbox{for} \quad n=1,3,5,7.
\end{equation}
$M_{1,9}(x,a)$ has the right power to be non-vanishing and we have
\begin{equation}
M_{1,9}(x,a) = {21\times 2^9\times P(a) \over y^6(x)} .
\end{equation}
For $n=11$ the resulting summation is also quite simple and we
have:
\begin{equation}
M_{1,11}(x,a) = {33\times 2^9\times P(a) \over y^6(x)} \times ( 6
\Delta_2(x) - \Delta^2_1(x) ) .
\end{equation}

For even $n$ we have the following results:
\begin{eqnarray}
M_{1,2}(x,a) & = & 0, \\
M_{1,4}(x,a) & = & {32 P(a) \, (x-a_1)^4 \over y^4(x)
\prod_{i=2}^6(a_1-a_i)} .
\end{eqnarray}
From the above results we see that although $M_{1,2n}(x,a)$ is not
modular invariant, it is invariant under a subgroup of the full
modular transformation. This subgroup of modular transformations
leaves $a_1$ fixed. This also explains why $M_{1,2}(x,a)$ is
vanishing because it should proportional to a homogeneous
polynomial $\tilde P(a) = \prod_{i<j= 2}^6 (a_i-a_j)$ which has
degree 10 while $M_{1,2}(x,a)$ is only a homogeneous polynomial of
degree 8 apart from the factor $y^4(x)$.

\section{The non-renormalization theorem}

For the non-renormalization theorem we need more identities. For
graviton and the antisymmetric tensor the vertex operator is (left
part only):
\begin{equation}
V_i(k_i,\epsilon_i, z_i) = ( \epsilon_i\cdot \partial X(z_i) + i
k_i\cdot \psi(z_i) \, \epsilon_i\cdot\psi(z_i) )\,  {\rm e}^{ i
k_i \cdot X(z_i, \bar z_i)} .
\end{equation}
By including the vertex operators we need to consider the
following extra spin structure dependent terms:
\begin{eqnarray}
\hbox{from}~{{\cal X}_1 + {\cal X}_6}: & & \langle
\psi(q_1)\psi(q_2) \prod_i k_i\cdot \psi(z_i) \,
\epsilon_i\cdot\psi(z_i) \rangle,
\label{eqpsi12} \\
\hbox{from}~{{\cal X}_2 + {\cal X}_3}: & & S_1(q)\, \langle
:\psi(q_1) \cdot\partial\psi(q_1) :\prod_i k_i\cdot \psi(z_i) \,
\epsilon_i\cdot\psi(z_i) \rangle . \label{eqpsi11}
\end{eqnarray}
The other terms are just the direct product of eq.~(\ref{eqform})
with the correlators from the vertex operators $\langle  \prod_i
k_i\cdot \psi(z_i) \, \epsilon\cdot\psi(z_i) \rangle$. Let's study
these direct product terms (may be called as disconnected terms)
first.

To prove the non-renormalization theorem we restrict our attention
to 3 or less particle amplitude. For the 3-particle amplitude we
have
\begin{equation}
\langle  \prod_{i=1}^3  k_i\cdot \psi(z_i) \,
\epsilon\cdot\psi(z_i) \rangle \propto
S(z_1,z_2)S(z_2,z_3)S(z_3,z_1) + \hbox{(other terms)}.
\end{equation}
By using the explicit expression of $S(z_1,z_2)$ we have
\begin{eqnarray}
S(z_1,z_2)S(z_2,z_3)S(z_3,z_1) & =  & { 1\over 8
z_{12}z_{23}z_{31}} \left\{ 2 + \left[{u(z_1) \over u(z_2)} +
{u(z_2) \over u(z_1)} \right] \right. \nonumber \\
& & \hskip -1.5cm  \left. + \left[{u(z_1) \over u(z_3)} + {u(z_3)
\over u(z_1)} \right] + \left[{u(z_2) \over u(z_3)} + {u(z_3)
\over u(z_2)} \right] \right\} \, .
\end{eqnarray}
These factors combined with the other factors in
eq.~(\ref{eqform}) give vanishing contribution to the $n$-particle
amplitude by using the following  ``vanishing identities":
\begin{eqnarray}
& & \sum_\delta \eta_\delta Q_\delta\left\{ {u(z_1) \over u(z_2)}
+ {u(z_2)\over u(z_1)} \right\}  \, S_n(x)  = 0, \qquad n = 1, 2, \\
& & \sum_\delta \eta_\delta Q_\delta \left\{ {u(z_1) \over u(z_2)}
- (-1)^n {u(z_2)\over u(z_1)} \right\}(S_1(x))^n = 0, \qquad n =
2, 3.
\end{eqnarray}
These identities can be proved by modular invariance and simple
``power counting" which we have explained in detail in the last
section.

Here we want to stress the importance of the limit $\tilde p_1 \to
q_{1,2}$. For arbitrary $\tilde p_{a}$, we would have a spin
structure dependent factor $S_1(q) (S_1(\tilde p_a))^2$ from
${\cal X}_{2,3}$ (specifically  from $T_\psi$, and other terms
from $T_{\beta\gamma}$ or  $\tilde B_{3/2}(\tilde p_a)$ are more
complicated as one can see from eqs.~(\ref{eq38}) and
(\ref{eq39})). So we need to compute the following expression:
\begin{equation}
\sum_\delta \eta_\delta Q_\delta\left\{ {u(z_1) \over u(z_2)} +
{u(z_2)\over u(z_1)} \right\}  \, S_1(q) \, (S_1(\tilde p_a))^2 .
\end{equation}
Unfortunately the above expression is not identically 0. We have:
\begin{eqnarray}
& & \sum_\delta \eta_\delta Q_\delta\, S_1(x) \left\{ {u(z_1)
\over u(z_2)} + {u(z_2)\over u(z_1)} \right\} \, \left[
\sum_{i=1}^3 (A_i - B_i)\right]^2 \nonumber \\
& & \hskip 3cm = { 8 P(a) (x-z_1)(x-z_2) \over y^2(x) y(z_1)
y(z_2) } \, (z_1-z_2)^2. \label{eq67}
\end{eqnarray}
Our conjecture is that the combined result would still be 0 and
independent of $\tilde p_a$'s. Nevertheless the above limit of
$\tilde p_1 \to q_{1,2}$ greatly simplifies the algebra in the
sense of making each term  to be 0 identically. This limit also
makes the computation of the four-amplitude doable (otherwise the
algebra would be much more complicated). Now we turn our attention
to the ``disconnected'' terms appearing in eqs.~(\ref{eqpsi12})
and (\ref{eqpsi11}).

The terms in eq.~(\ref{eqpsi12}) have already been discussed in
\cite{IengoZhu1}. Here we briefly review the argument. We have
\begin{equation}
\langle \psi(q_1)\psi(q_2) \prod_i k_i\cdot \psi(z_i) \,
\epsilon\cdot\psi(z_i) \rangle \propto S(q_1,z_1)
S(z_1,z_2)S(z_2,z_3) S(z_3,q_2)  + \cdots .
\end{equation}
By using the explicit expression of $S(z,w)$ and note that $u(q_2)
= - u(q_1)$ we have
\begin{eqnarray}
& & S(q_1,z_1) S(z_1,z_2)S(z_2,z_3) S(z_3,q_2) \propto
\sum_{i=1}^3 \left[ {u(q_1)\over u(z_i)} - {u(z_i)
\over u(q_1)} \right]  \nonumber \\
& & \qquad + \sum_{i<j=1}^3 \left[{u(z_i)\over u(z_j)} - {u(z_j)
\over u(z_i)} \right]   + {u(q_1) u(z_2)\over u(z_1) u(z_3)}
 - {u(z_1) u(z_3) \over u(q_1) u(z_2)} .
\end{eqnarray}
These terms also give  vanishing  contributions as we can prove
the following identities:
\begin{eqnarray}
& & \sum_\delta \eta_\delta Q_\delta\left\{ {u(z_1) \over u(z_2)}
- {u(z_2)\over u(z_1)} \right\}     = 0,   \label{eq70} \\
& & \sum_\delta \eta_\delta Q_\delta \left\{ {u(z_1) u(z_2) \over
u(z_3) u(z_4)} -   {u(z_3) u(z_4)\over u(z_1) u(z_2)} \right\}  =
0.
\end{eqnarray}
These identities were firstly proved in \cite{IengoZhu1}. The
proof is quite simple by using modular invariance. For example we
have
\begin{eqnarray}
& & \sum_\delta \eta_\delta Q_\delta\left\{ {u(z_1) \over u(z_2)}
- {u(z_2)\over u(z_1)} \right\}
\nonumber \\
& & = { 1\over y(z_1) y(z_2) } \sum_\delta \eta_\delta
Q_\delta\left\{ \prod_{i=1}^3(z_1-A_i)(z_2-B_i) -
\prod_{i=1}^3(z_1-B_i)(z_2-A_i) \right\}
\nonumber \\
& & \propto {(z_1 - z_2)\, P(a) \over y(z_1)y(z_2)} ,
\end{eqnarray}
which must be vanishing as the degrees of the  homogeneous
polynomials (in $a_i$ and $z_j$) don't match. Here we have  used
again the modular invariance of the above expression.\footnote{The
minus sign in eq.~(\ref{eq70}) makes the expression invariant
under the all the modular transformations. With a plus sign the
expression is only invariant under a subgroup of the modular
transformation. Nevertheless eq.~(\ref{eq70}) is still true with a
plus sign. The explicit results are:
\begin{eqnarray}
& & \sum_\delta \eta_\delta Q_\delta\left\{ {u(z_1) \over u(z_2)}
+ {u(z_2)\over u(z_1)} \right\}     = 0,   \\
& & \sum_\delta \eta_\delta Q_\delta \left\{ {u(z_1) u(z_2) \over
u(z_3) u(z_4)} + {u(z_3) u(z_4)\over u(z_1) u(z_2)} \right\}  =
{ 2 P(a) z_{13}z_{14}z_{23}z_{24}\prod_{i=1}^4(a_1 -z_i) \over
 \prod_{i=1}^4y(z_i) \prod_{i=2}^6(a_1-a_i)} .
\end{eqnarray} }

The last term we need to compute is the term in eq.
(\ref{eqpsi11}). We have
\begin{eqnarray}
& & \langle :\psi(q_1) \cdot\partial\psi(q_1) : \prod_i k_i\cdot
\psi(z_i) \, \epsilon\cdot\psi(z_i) \rangle_c
   =  K(1,2,3) \nonumber \\
& &  \qquad  \times (S(q_1,z_1,z_2,z_3) + S(q_1,z_2,z_3,z_1) +
S(q_1,z_3,z_1,z_2)
 \nonumber \\
& &  \qquad   - S(q_1,z_1,z_3,z_2) - S(q_1,z_2,z_1,z_3) -
S(q_1,z_3,z_2,z_1) ), \label{eqpsi33}
\end{eqnarray}
where
\begin{eqnarray}
K(1,2,3) & = &
k_1\cdot\epsilon_3 k_2\cdot\epsilon_1 k_3\cdot\epsilon_2-
k_1\cdot\epsilon_2 k_2\cdot\epsilon_3 k_3\cdot\epsilon_1
\nonumber \\
& & + k_1\cdot k_2(k_3\cdot\epsilon_1\epsilon_2\cdot\epsilon_3 -
k_3\cdot\epsilon_2\epsilon_1\cdot\epsilon_3) \nonumber \\
& & + k_2\cdot k_3(k_1\cdot\epsilon_2\epsilon_3\cdot\epsilon_1-
k_1\cdot\epsilon_3\epsilon_2\cdot\epsilon_1) \nonumber \\
& & + k_3\cdot k_1(k_2\cdot\epsilon_3\epsilon_1\cdot\epsilon_2-
k_2\cdot\epsilon_1\epsilon_3\cdot\epsilon_2)   \, , \\
S(x,z_1,z_2,z_3) & = & S(x,z_1) S(z_1,z_2)
S(z_2,z_3)\partial_xS(z_3,x) .
\end{eqnarray}
We note that $K(1,2,3)$ is invariant under the cyclicly
permutations of (1,2,3). It is antisymmetric under the interchange
$2 \leftrightarrow 3$. We have used these properties in eq.
(\ref{eqpsi33}).

To compute explicitly these expressions we first note the following:
\begin{equation}
\partial_x S(z, x) = { 1\over  2(z-x)^2 }
\, {u(z) + u(x) \over \sqrt{u(z)u(x)}  }  - {S_1(x) \over 8 \,
(z-x) } \, {u(z) - u(x) \over \sqrt{u(z)u(x)} } \,   .
\end{equation}
In order to do the summation over spin structure  we need a
``non-vanishing identity". This and other identities needed in the
4-particle amplitude computations are summarized as follows:
\begin{eqnarray}
& &  \sum_\delta \eta_\delta Q_\delta \left\{ {u(z_1)  u(z_2)
\over u(z_3)u(z_4)} - (-1)^n {u(z_1)u(z_2)\over u(z_3)u(z_4) }
\right\} (S_m(x))^n \nonumber  \\
& &  \qquad \qquad =  {2 P(a) \prod_{i=1}^2\prod_{j=3}^4 (z_i-z_j)
\prod_{i=1}^4(x-z_i) \over y^2(x) \prod_{i=1}^4 y(z_i) } \times
C_{n,m},
\end{eqnarray}
where
\begin{eqnarray}
C_{1,1} & = & 1,  \label{eq991} \\
C_{2,1} & = & - 2 (\tilde z_1 + \tilde z_2 - \tilde z_3 - \tilde
z_4) ,     \\
C_{1,2} & = & \Delta_1(x)  - \sum_{k=1}^4 \tilde z_k   , \\
C_{3,1} & = & 2 \Delta_2(x) - \Delta_1^2(x) + 2 \Delta_1(x)\,
\sum_{k=1}^4 \tilde z_k \nonumber \\
& &  + 4 \sum_{k<l} \tilde z_k \tilde z_l   - 12 ( \tilde z_1  +
\tilde z_2 )(\tilde z_3 + \tilde z_4 ) \, , \label{eq992} \\
\tilde z_k & = & { 1\over x - z_k}, \\
 P(a) & = & \prod_{i<j}(a_i-a_j).
\end{eqnarray}
$C_{1,1}$ and $C_{1,2}$ were derived in \cite{IengoZhu2}. Although
other values of $n,m$ also gives modular invariant expressions,
the results are quite complex.\footnote{This is due to the
non-vanishing of the summation over spin structures when we set
$z_1=z_3$ or $z_1=z_4$, etc.} Fortunately we only need to use the
above listed results. The proof of these summation formulas will
be given in \cite{AllZhu3}.

By using these formulas we have:
\begin{equation}
\sum_\delta \eta_\delta Q_\delta
S(x,z_1,z_2,z_3) \,  S_1(x)  = - {P(a)\over16  y^2(x) } \prod_{i=1}^3
{x-z_i\over y(z_i)} .
\end{equation}
We note that the above formula is invariant under the interchange
$z_i \leftrightarrow z_j$.

By using this result and   eq.~(\ref{eqpsi33}), we have:
\begin{equation}
\sum_\delta \eta_\delta Q_\delta S(q_1,q_2) \langle
:\psi(q_1)\cdot\partial\psi(q_1):\prod_{i=1}^3 k_i\cdot\psi(z_i)
\epsilon_i\cdot\psi(z_i)\rangle_\delta =0.
\end{equation}
This completes our verification of the non-renormalization theorem
at two loops.



\section*{Acknowledgments}

Chuan-Jie Zhu would like to thank Roberto Iengo for reading the
paper and comments. He would also like to thank E. D'Hoker and D.
Phong for discussions and Jian-Xin Lu and the hospitality at the
Interdisciplinary Center for Theoretical Study, Physics,
University of Science and Technology of China.

\end{document}